\documentclass[%
 aip,
 jmp,%
 amsmath,amssymb,
 reprint,%
unsortedaddress
]{revtex4-1}

\usepackage{graphicx}
\usepackage{dcolumn}
\usepackage{bm}
\usepackage{float}
\usepackage{gensymb}
\usepackage{color}

\begin{document}
	\title{Interaction of Fluorescent Gold Nanoclusters with Transition Metal Dichalcogenides Nanosheets: A Spectroscopic Study}

		\author{Arun Singh Patel} 
		\affiliation{School of Computational \& Integrative Sciences, Jawaharlal Nehru University, New Delhi-110067, India}
		\affiliation{Department of Physics, Hindu College, University of Delhi, New Delhi-110007}
		\email{arunspatel.jnu@gmail.com}
	
	\author{Anirban Chakraborti} 	
	\affiliation{School of Computational \& Integrative Sciences, Jawaharlal Nehru University, New Delhi-110067, India}
	\affiliation{Centro Internacional de Ciencias, Cuernavaca-62210, M\'{e}xico}
	\email{anirban@jnu.ac.in}
	
			\author{Praveen Mishra} 
			\affiliation{School of Computational \& Integrative Sciences, Jawaharlal Nehru University, New Delhi-110067, India}


	\begin{abstract}
	In this paper, the interaction of few tens of atoms containing gold nanoclusters with two dimensional nanosheets of transition metal dichalcogenides nanosheets has been explored. The gold nanoclusters have been synthesized using chemical reduction method in presence of protein molecules as stabilizing agent. The transition metal dichalcogenides nanosheets of molybdenum disulfide (MoS$_2$) has been chemically exfoliated. Different microscopic and optical spectroscopic tools have been used for characterizing the physical properties of the gold nanoclusters and the two dimensional nanosheets of MoS$_2$. The gold nanoclusters exhibit fluorescence emission at 690 nm. However, the interaction with transition metal dichalcogenides diminishes drastically the fluorescence intensity of the nanoclusters. The spectroscopic methods used for understanding the interaction in the system reveals the absence of energy transfer and dynamic nature of the fluorescence quenching.       	
		
	\end{abstract}
	
	
	\maketitle
	
	\section{Introduction}
In recent years, the metal nanoclusters (NCs) consisting of few tens of metal atoms have been studied extensively due to their intriguing  physical properties and excellent chemical stability. \cite{yarramala2017, lian2017, hu2018, abbas2018, pyo2015} The size of these nanoclusters  lies in between individual atoms and nanoparticles. These metal nanoclusters provide a bridge between atomic and nanoparticles behavior  of noble metals. \cite{ pateljmc2014,pateljnn2016} At this length scale, the size of nanoclusters is comparable to the Fermi wavelength of the free electrons of the metal, which  results in quantum confinement of the free electrons in the nanoclusters. The quantum confinement leads to widening of the energy gap between  the discrete electronic energy levels.  The discrete energy levels enable these nanoclusters to act as highly  optically  active materials in the visible range, which is generally  absent in the case of nanoparticles and the bulk metal. 
Similar to the fluorescent dye molecules, gold nanoclusters also show strong fluorescence at low temperature as compared to high temperature.  The relationship among the energy of emitted photons ($E_g$) from metal nanoclusters  and  number of atoms ($N$) and the Fermi energy  ($E_F$) of metal can be  represented in the following form \cite{zheng2007}

\begin{equation}
	E_g=\frac{E_F}{N^{1/3}}.
\end{equation}

Gold nanoclusters are less toxic as compared to other metal nanoclusters. Hence, gold nanoclusters can be a good choice as fluorescent materials in comparison with available fluorescent dye molecules. The dye molecules that are used as fluorescent tags in spectroscopic analysis have toxic nature; so, they have limitations in direct use in biological studies. In recent years, there has been great interest among the researchers to understand the interaction of fluorescent materials with different kinds of nanomaterials, starting from dye molecules with quantum dots to quantum dots with 2D nanomaterials.\cite{dong2010, clapp2004, hohng2005, prasai2015, goodfellow2016, prins2014} These interactions have been explored for different areas of application like sensing, solar cell, imaging, etc. Since discovery of graphene, people started searching for other forms of 2D nanosheets and in recent times different 2D nanomaterials, which have varying band gap from UV to IR region, are being extensively studied.  Among these 2D nanosheets, the nanosheets of transition metal dichalcogenides (TMDCs), which have band gap in the visible wavelength range found great interest. \cite{chhowalla2013, wang2012, guan2015, wu2018, luo2018}  These TMDCs possess indirect band gap when they are in bulk form and show direct band gap while in single layer form.\cite{arunapl2016, lee2012, chakrabortipla2016} 
Among these TMDCs nanosheets, molybdenum disulfide MoS$_2$ and tungsten disulfide WS$_2$ have been studied extensively.  The nanosheets of MoS$_2$  have number of layers dependent bad gap and hence the band gap can be tuned easily.   This attracted tremendous attention due to its intriguing chemical, physical and mechanical properties. Good biocompatibility, flexible surface and excellent fluorescence quenching capability have made MoS$_2$ compatible for numerous applications. \cite{srinivasan2018, paredes2016, lee2013} Understanding the interaction between MoS$_2$ and gold nanoclusters is the key factor for the development of biosensors, drug delivery and therapeutic systems. In this paper, the interaction of gold nanoclusters with MoS$_2$ nanosheets has been investigated. This work is important for exploring the possibilities of energy transfer in nanomaterials where biomolecules are present at the interface of different nanomaterials, which can be further explored for sensing applications. 

\section{Methods and Results}
\textit{Synthesis of gold nanoclusters:} For synthesis of gold nanoclusters, chloroauric acid (HAuCl$_4$.3H$_2$O, Sigma Aldrich) was used as precursor of gold nanoclusters. 
 Gold nanoclusters were synthesized using  chemical reduction method. In typical synthesis procedure, 50 mg bovine serum albumin (BSA) was dissolved in 1 mL of distilled water. In the BSA solution, 1 mL of 10$^{-2}$ M gold salt aqueous solution was added, and the mixture was stirred for 5 min. Further, 0.1 mL aqueous solution of 1M sodium hydroxide was injected into the  solution, and the mixture was allowed to be stirred for 10 min. As-prepared solution was incubated at 37 $^0$C for 24 h. The yellow colored solution turned to a reddish-brown after formation of gold nanoclusters. The aqueous dispersion of gold nanoclusters was kept at 4  $^0$C for further study.

\textit{Chemical exfoliation of MoS$_2$ nanosheets:} The MoS$_2$ nanosheets were obtained by chemical exfoliation of the bulk MoS$_2$ powder using ultrasonication technique. For this purpose, a stock solution of BSA was prepared (1 mg/mL). The solution was used as solvent for dispersion of bulk MoS$_2$ powder. 50 mg of MoS$_2$ powder was dispersed in 10 mL of the aqueous solution of BSA. The dispersion of MoS$_2$ was ultrasonicated for 20 h. After ultrasonication, the precipitate was allowed to settle down and the supernatant was collected, and a stock solution of MoS$_2$ was made.  

\textit{Characterization techniques:} The gold nanoclusters and MoS$_2$ nanosheets were characterized by different spectroscopic and microscopic techniques. The absorption spectra of these nanomaterials were recorded using UV-visible spectrophotometer (IMPLEN, Inkarp). The Raman spectrometer from WITEC was used to investigate the Raman active vibrational modes of MoS$_2$.  For this purpose, Raman spectrometer, with 532 nm laser excitation source was used. The shape and size of the nanosheets were analyzed by transmission electron microscope (JEOL-2010) operated at 200 keV.  The fluorescence spectra of gold nanoclusters in absence and presence of MoS$_2$ nanosheets were recorded using fluorimeter. The fluorescence lifetime decay was monitored using time correlated single photon counting set-up (FL920, Edinburgh Instruments, UK). The excitation source was 370 nm pulse laser with operating frequency of 20 MHz.

\begin{figure}
	\centering
	\includegraphics[width=0.95\linewidth]{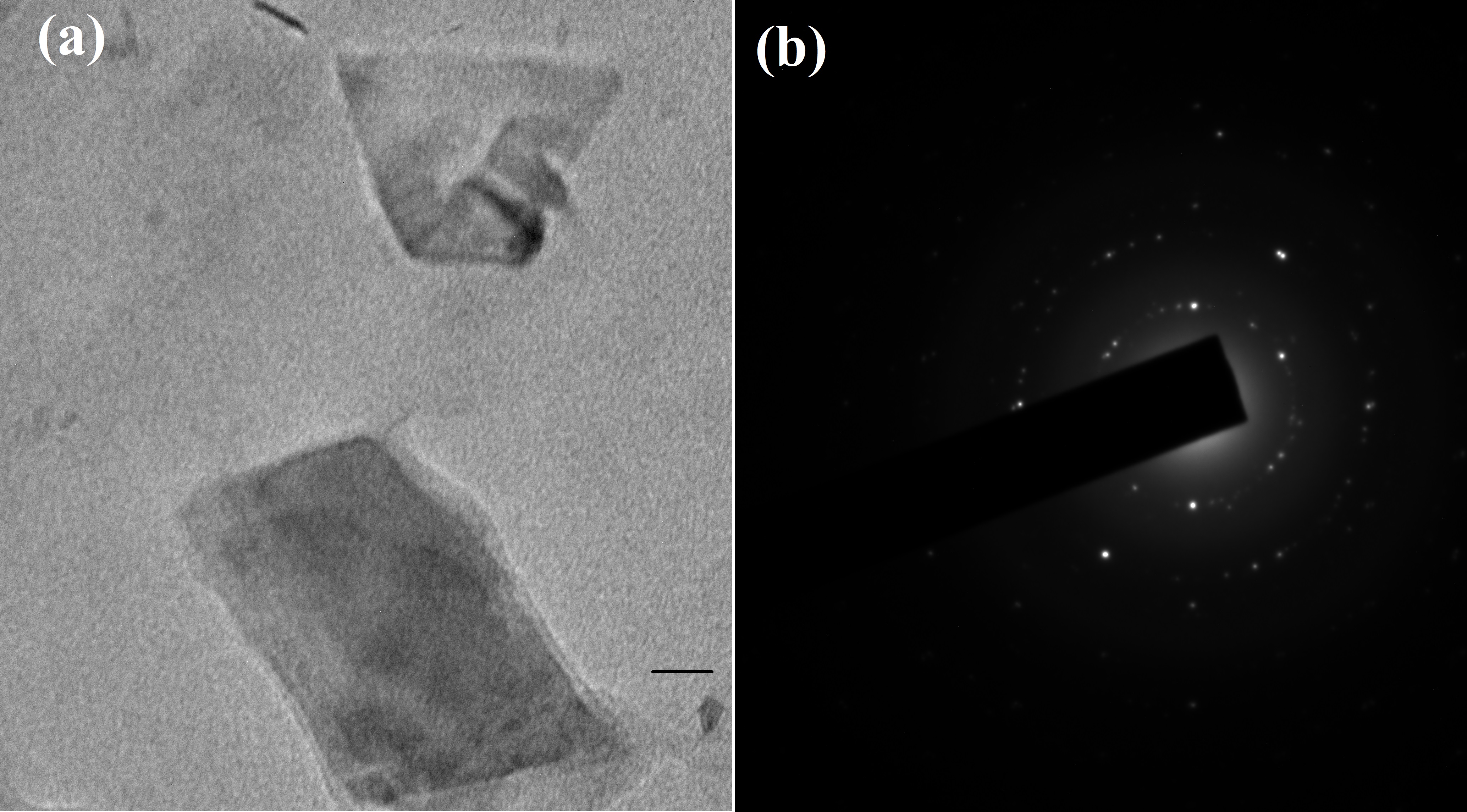} 
	\caption{TEM images  of (a) MoS$_2$ nanosheets, and (b) the selective area electron diffraction pattern of the nanosheets,  scale bar equals to 20 nm. }
	\label{TEM}
\end{figure}

The shape and morphology of  MoS$_2$ nanosheets were monitored by TEM analysis. The TEM image of MoS$_2$ nanosheets is shown in Figure \ref{TEM}, where sheet like structure of MoS$_2$ is observed. 
A  color contrast in the TEM image is observed in Figure \ref{TEM}(a), which arises due to few layers of the nanosheets.  Figure \ref{TEM}(b) is the selective electron diffraction (SAED) pattern, where hexagonal spots are observed due to the hexagonal lattice structure of MoS$_2$.   

\begin{figure}
	\centering
	\includegraphics[width=0.95\linewidth]{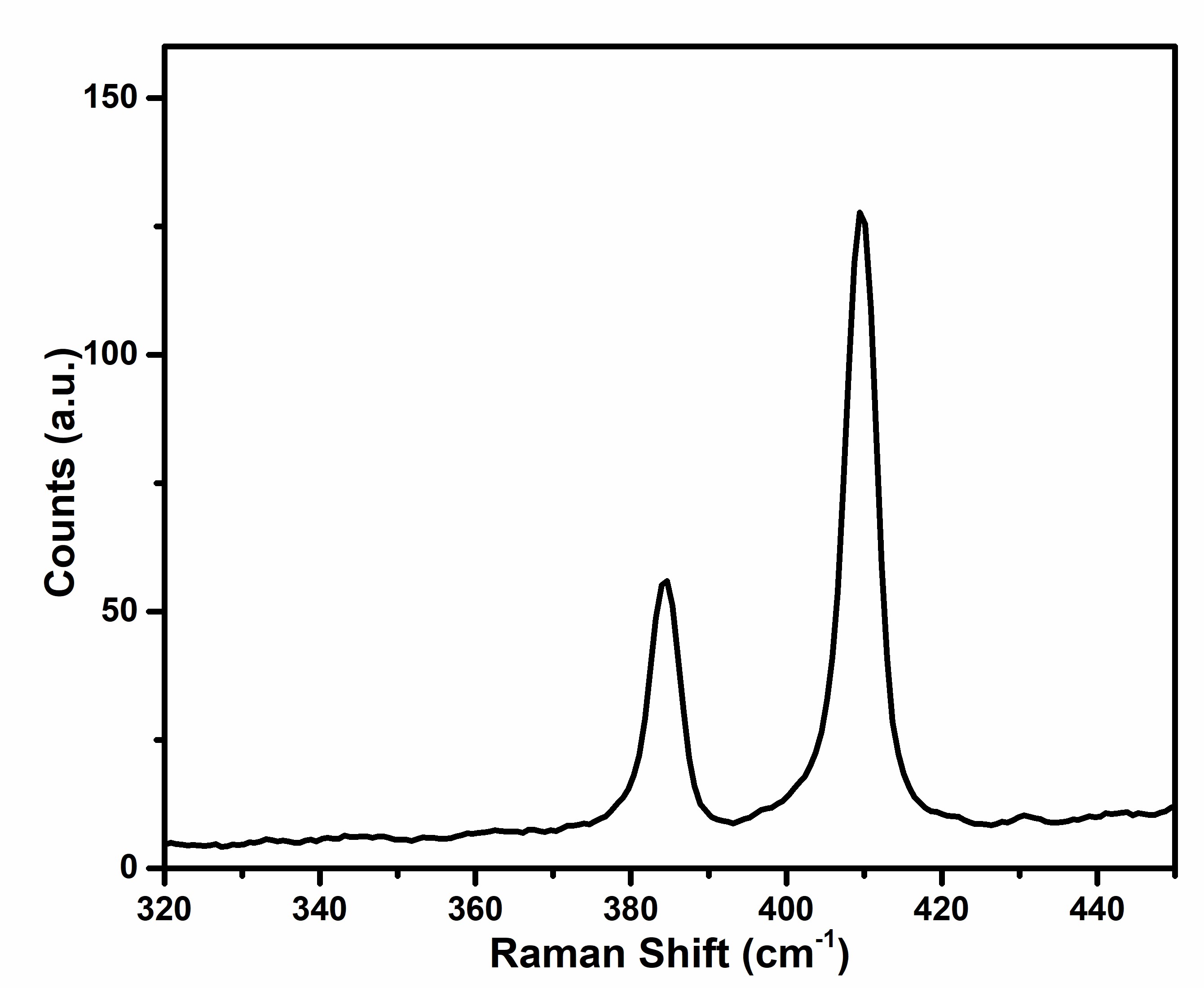} 
	\caption{Raman spectrum of MoS$_2$ nanosheets.  }
	\label{Raman}
\end{figure}

The Raman spectrum of MoS$_2$ nanosheets is shown in Figure \ref{Raman}. In case of MoS$_2$ nanosheets, there are two distinct Raman peaks, which are observed at 385 and 409 cm$^{-1}$ and these peaks are assigned as $E^1_{2g}$ and $A_{1g}$ vibrational modes, respectively. 
The $E^1_{2g}$ peak originates due to in-plane vibration of the sulfur and molybdenum atoms in the lattice, while the $A_{1g}$ peak is due to out-of-plane vibration of the sulfur atoms.  

\begin{figure}
	\centering
	\includegraphics[width=0.95\linewidth]{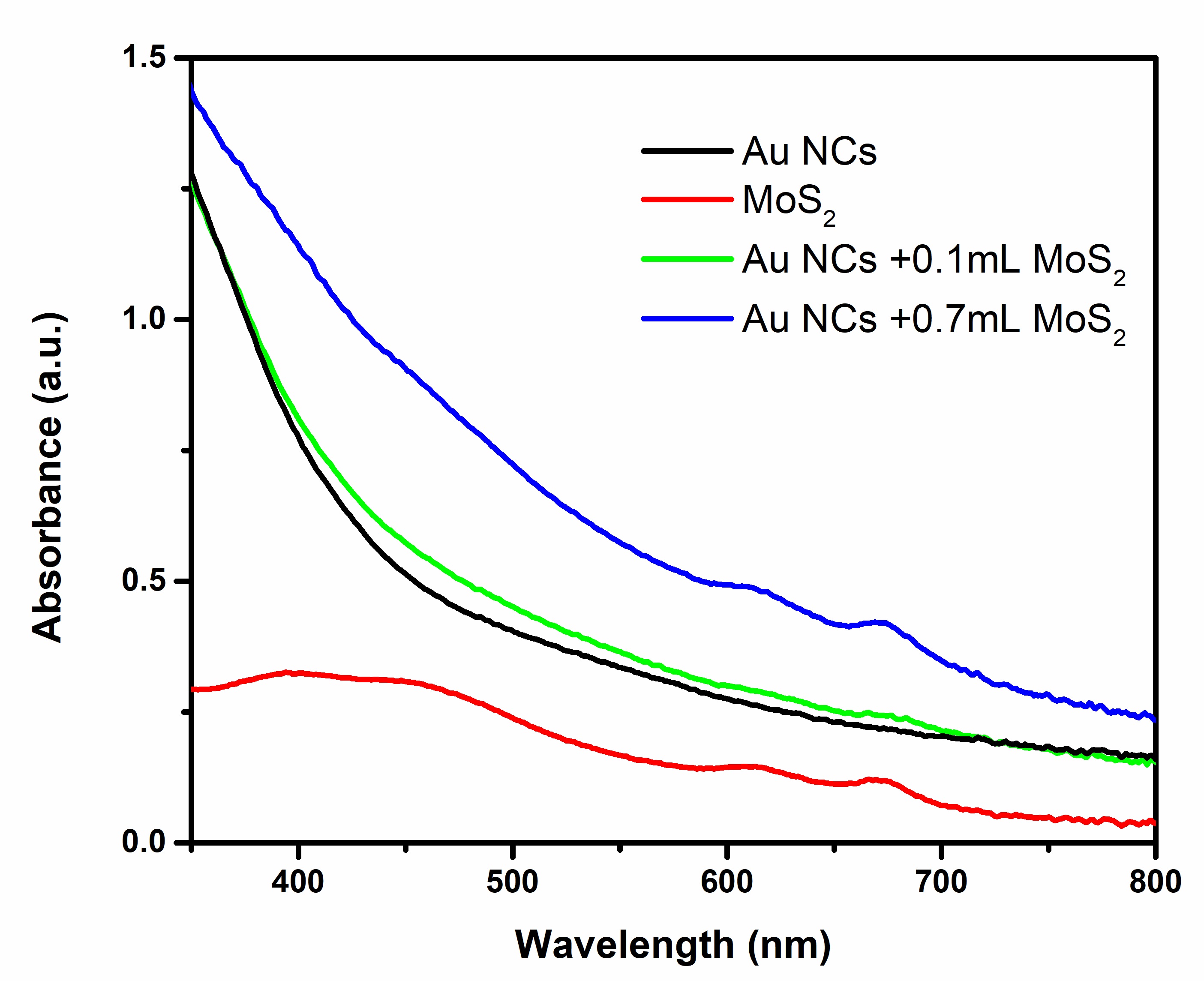} 
	\caption{Absorption spectra of gold nanoclusters, MoS$_2$ nanosheets, and mixed dispersion of gold nanoclusters and MoS$_2$ nanosheets.   }
	 \label{Abs}
\end{figure}

The optical absorption of gold nanoclusters and MoS$_2$ nanosheets are recorded using absorption spectrophotometer in the visible range. The absorption spectra of gold nanoclusters, MoS$_2$ and mixture of both are shown in Figure \ref{Abs}.   
The gold nanoclusters have continuously increasing absorbance towards the lower wavelength of the spectrum. The absorption spectrum of MoS$_2$ exhibits two distinct absorption peaks -- one appearing at 613 nm and the other at 670 nm, which are known as B and A excitonic peaks, respectively. 

\begin{figure}
	\centering
	\includegraphics[width=0.95\linewidth]{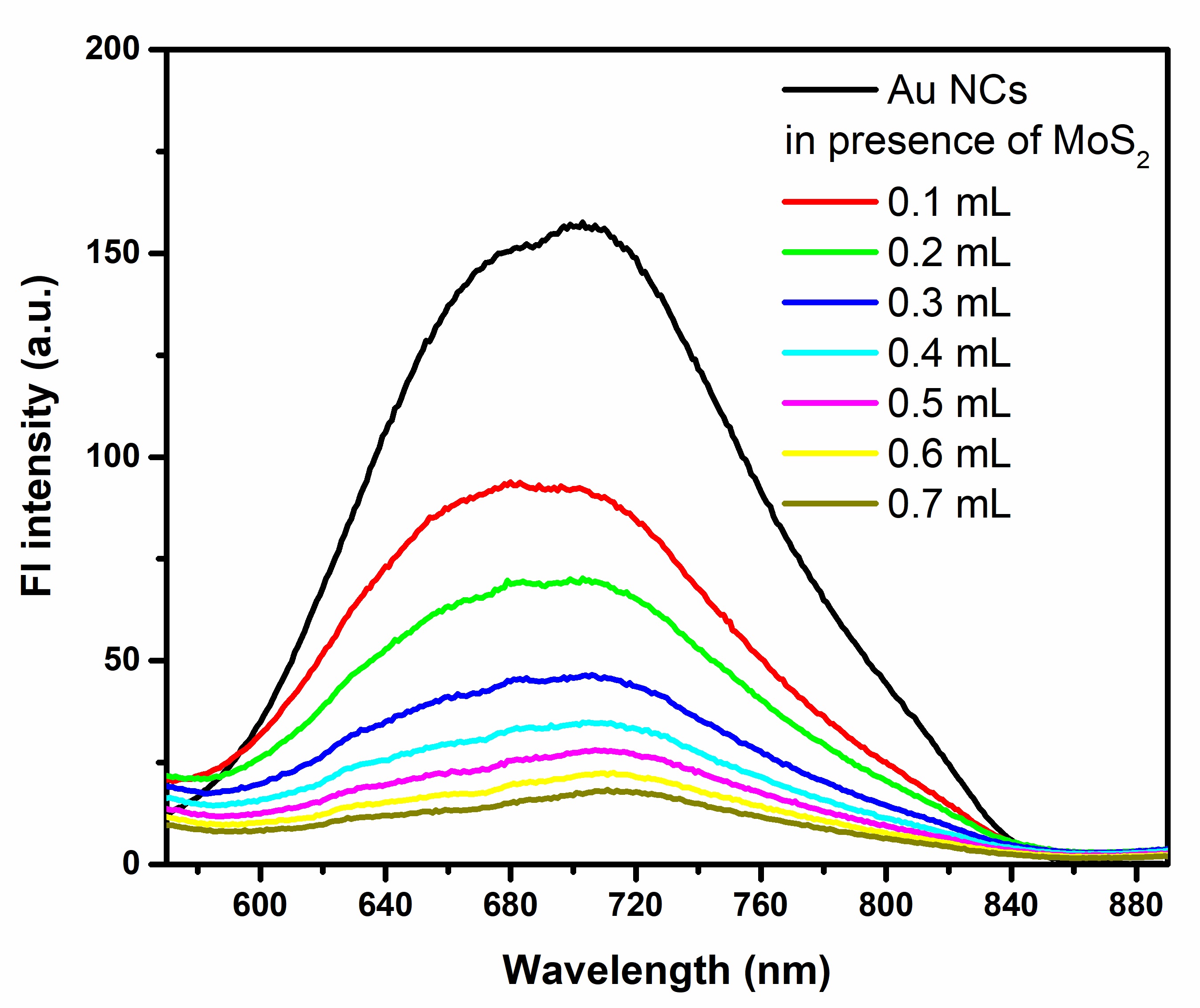} 
	\caption{Fluorescence spectra of gold nanoclusters in absence as well as in the presence of different quantity of  MoS$_2$ nanosheets.    }
	\label{FL_MoS2}
\end{figure}

\begin{figure}
 	\centering
 	\includegraphics[width=0.95\linewidth]{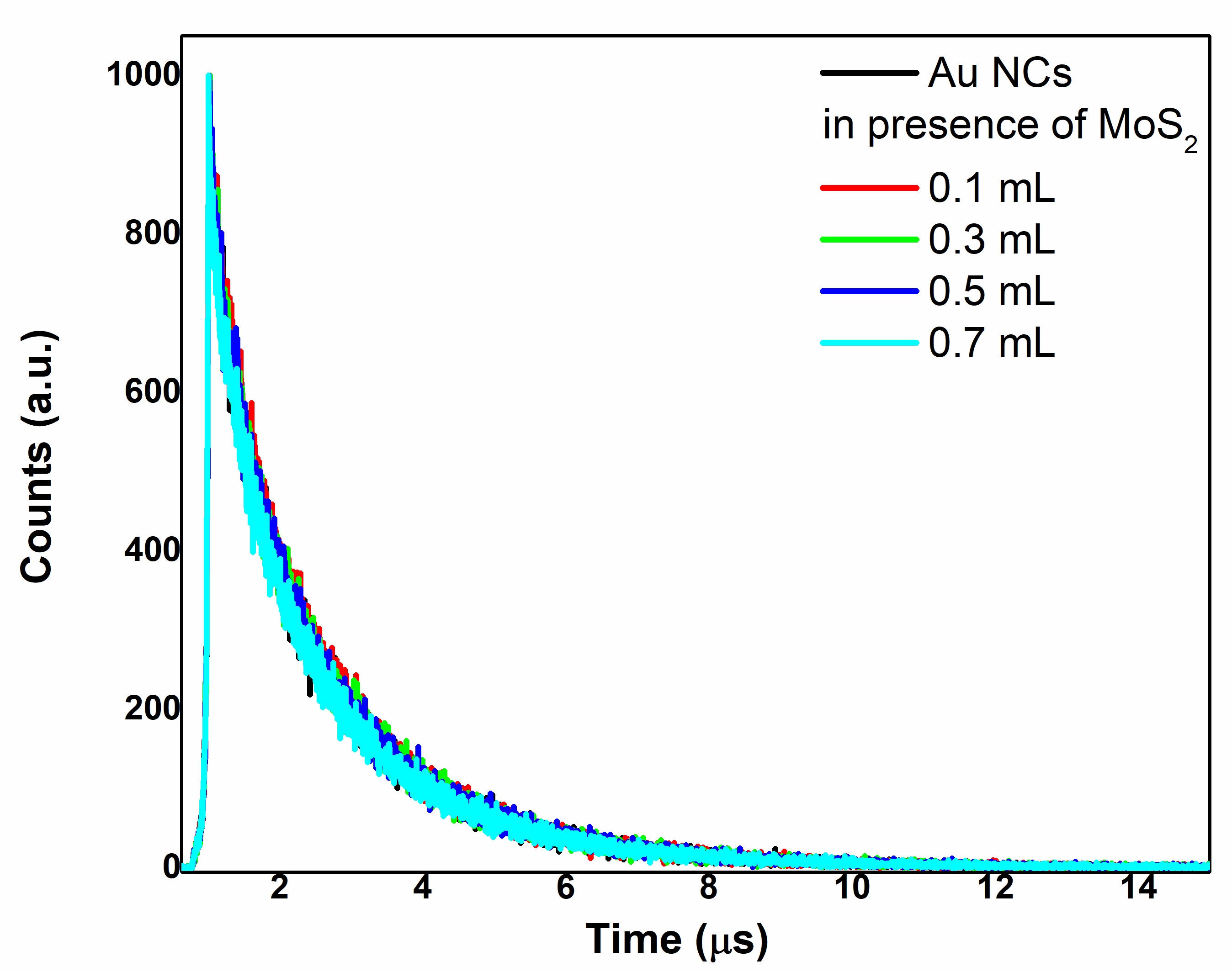} 
 	\caption{Fluorescence lifetime decay curves of gold nanoclusters in absence and presence of different quantity of MoS$_2$ nanosheets. }
 	\label{TRFS}
 \end{figure}

In order to investigate the effect of the presence of MoS$_2$ on the fluorescence property of the gold nanoclusters, the fluorescence spectra of gold nanoclusters  have been recorded, both in absence and presence of different quantities of MoS$_2$ nanosheets. 
The fluorescence spectra of gold nanoclusters in the absence and presence of MoS$_2$ are shown in Figure \ref{FL_MoS2}. 
The fluorescence (Fl) emission of gold nanoclusters is observed at 690 nm. In presence of MoS$_2$ nanosheets the Fl intensity of gold nanoclusters diminishes drastically. The Fl quenching may be due to either static or dynamic nature of quenching. \cite{pateljfl2016, cheng2006} Using  equation \ref{eq_energytime}, the Fl quenching efficiency was estimated
     \begin{equation}
     	E=1-\frac{F_{DA}}{F_D} \label{eq_energytime},
     \end{equation}
where F$_{DA}$ and F$_D$ are fluorescence intensities of gold nanoclusters in presence and absence of MoS$_2$ nanosheets. The Fl quenching efficiency was found to be 41 and 89 \% for MoS$_2$ nanosheets with 0.1 mL and 0.7 mL of the acceptor quantities, respectively.  Further, the nature of the fluorescence quenching was investigated using time resolved fluorescence spectroscopic (TRFS) technique. The fluorescence lifetime of the gold nanoclusters was estimated from the fluorescence lifetime decay curve of the gold nanoclusters. The fluorescence lifetime decay curves of gold nanoclusters in absence and presence of the nanosheets, are shown in Figure \ref{TRFS}. 
The fluorescence lifetime decay curves were fitted with exponential decaying function of the following form \cite{patelapl2014}
  \begin{equation}
  I(t)=\sum\limits_{j=1}^{m}\alpha_j \exp(-t/\tau_j), \label{eq_expfit}
  \end{equation}
 where  $\alpha_j$ is the weighing factor, $\tau_j$ is the fluorescence lifetime associated with $j^{th}$ component. For gold nanoclusters, the Fl decay curves were fitted with triple exponential decaying functions. The three components are associated with three different environments of the emitters. In gold nanoclusters, it has been observed that there are few gold atoms  in Au(I) state, while the emission comes out due to Au(0) atoms, which are forming core of the gold nanoclusters. The gold nanoclusters are formed  with 25 gold atoms, which are in conjugation with BSA molecules. Few gold atoms bind with sulfur atoms of the BSA molecules and thus the inner gold atoms are in Au(0) state. Thus, one component is arising due to core and the other due to outer interacting nanoclusters. The outer component is responsible for high Fl lifetime value, while the inner one is responsible for small value of the Fl lifetime. 
The average Fl lifetime of gold nanoclusters is found to be 1.1  $\mu$s. From Figure \ref{TRFS}, it is observed that the nature of the decaying curves remains the same in presence of MoS$_2$ nanosheets, even at very high concentrations of the MoS$_2$ nanosheets. Thus,  the Fl lifetime remains unaltered in presence of MoS$_2$ nanosheets.   
 
\begin{figure}
	\centering
	\includegraphics[width=0.95\linewidth]{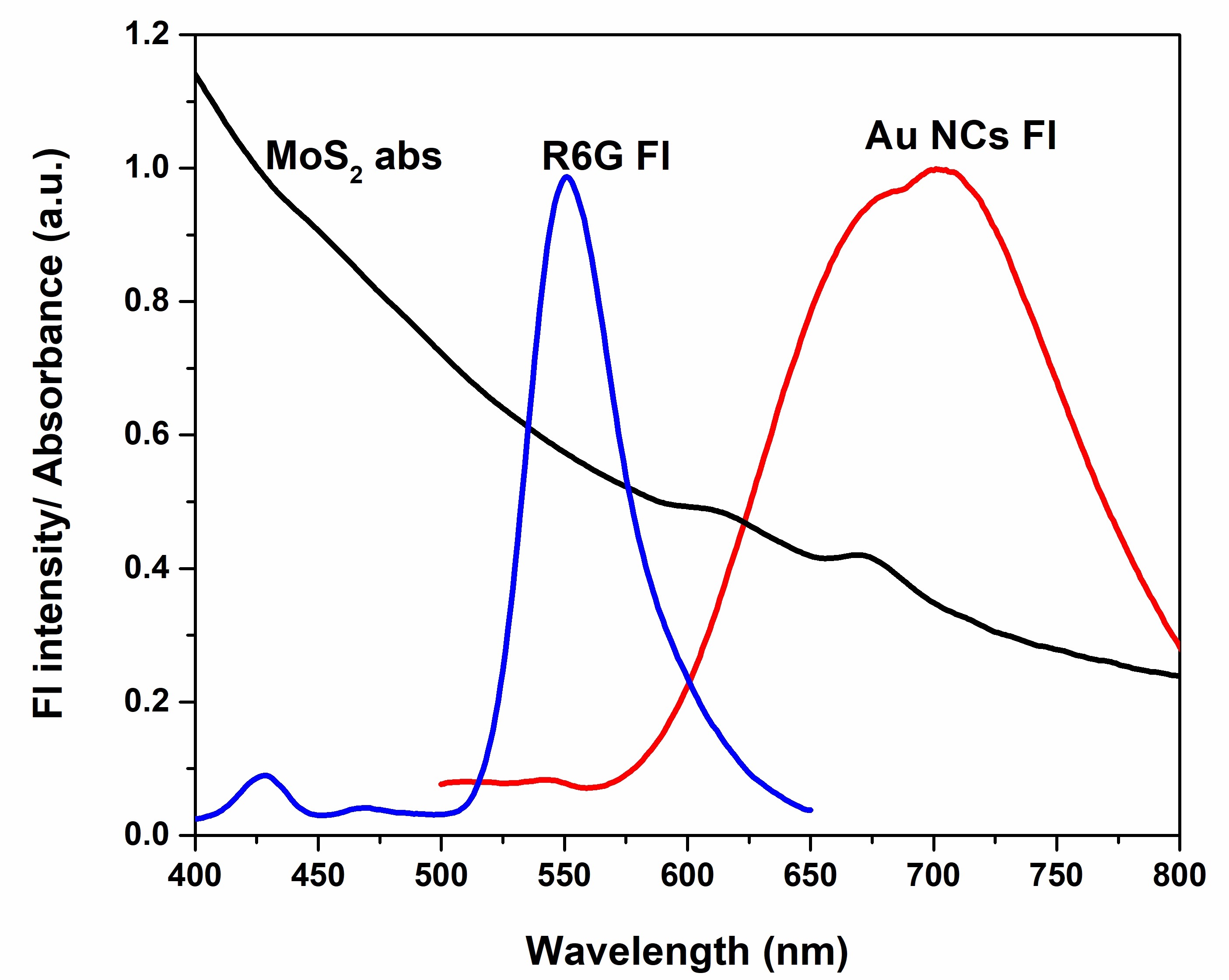} 
	\caption{Absorption spectrum of MoS$_2$ nanosheets and emission spectra of gold nanoclusters and R6G molecules showing the overlapping of absorption and emission spectra.      \label{Overlap}}
\end{figure}
\begin{figure}
	\centering
	\includegraphics[width=0.95\linewidth]{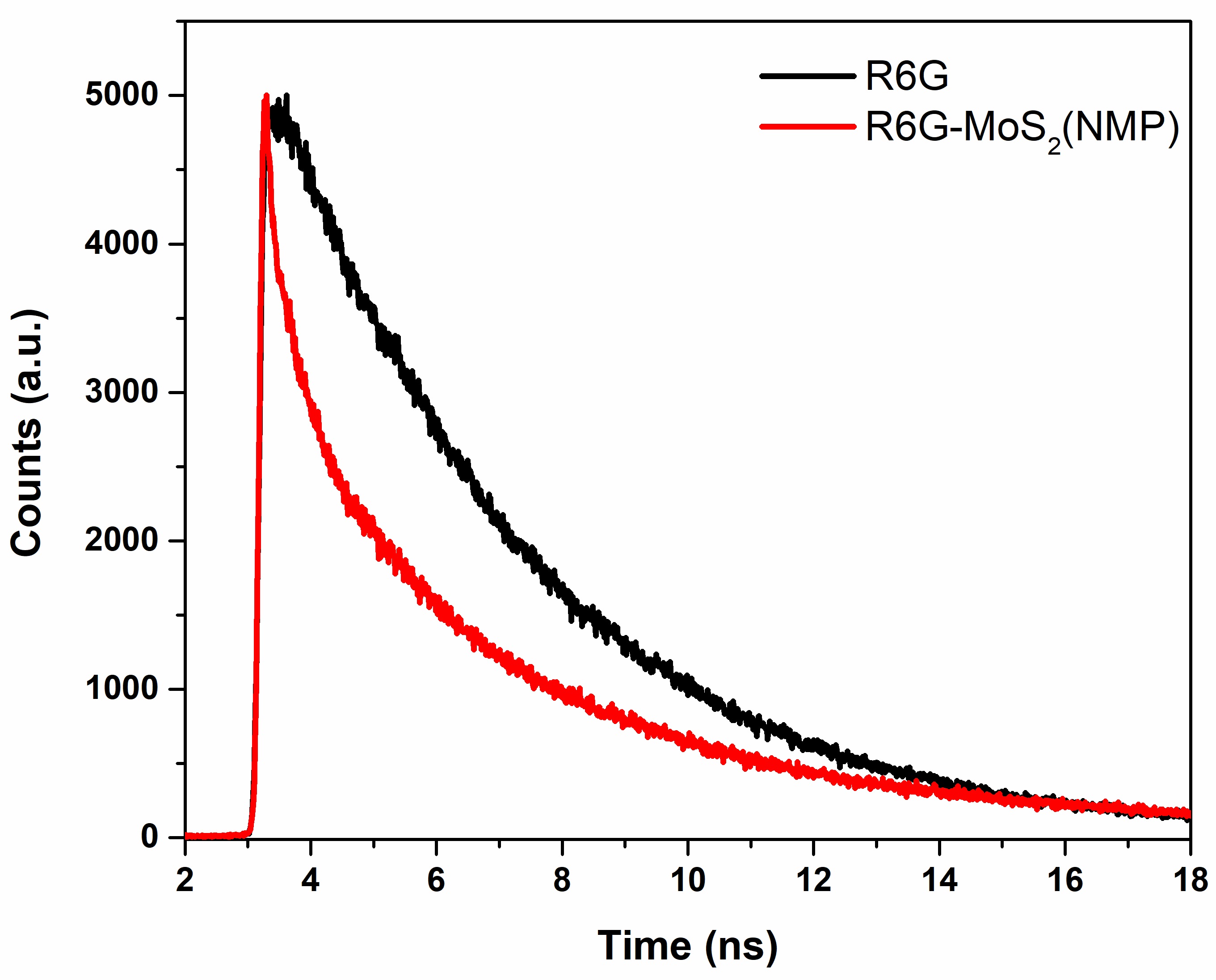} 
	\caption{Fluorescence lifetime decay curves of R6G molecules in absence and presence of MoS$_2$ nanosheets, here MoS$_2$ nanosheets are exfoliated in NMP solvent. \label{TRFS-NMP}}
\end{figure}

The time resolved fluorescence  study shows that there is no change in the fluorescence lifetime of gold nanoclusters in presence of MoS$_2$ nanosheets even at very high concentration of the nanosheets. Previously, it had been reported that the energy transfer from adjacent BSA molecules to the MoS$_2$ nanosheets could possibly occur.\cite{patelrsc2017} In that case, the distance between donor and acceptor must be within the range of resonance energy transfer phenomenon. However in the present case, the gold nanoclusters are synthesized in conjugation with the BSA molecules separately, and the BSA molecules are used for chemically exfoliation of the MoS$_2$ nanosheets. This causes the enlargement of the distance between the gold nanoclusters and the MoS$_2$ nanosheets, which limits the possibility of any kind of energy transfer, even though there is a  overlap of the emission spectrum of gold nanoclusters and the absorption spectrum of the MoS$_2$ nanosheets (see Figure \ref{Overlap}). There is a complete overlap of A excitonic peak of MoS$_2$ with the Fl emission peak of the gold nanoclusters, and partial overlap of the B excitonic peak. In the present case, the distance between the emitters and the acceptors is beyond the limit of the fluorescence resonance energy transfer, which is typically in the range of 1-10 nm. In case of gold nanoclusters and MoS$_2$ nanosheets, the intermediate layer of the BSA molecules helps to increase the distance between emitters and acceptors beyond the permitted upper limit of the distance. This was further confirmed by using fluorescent dye molecules, Rhodamine 6G (R6G) as donor molecules in place of Au NCs.  R6G molecules show Fl emission around 550 nm, which has an overlap with the absorption spectrum of MoS$_2$ (shown in Figure \ref{Overlap}).   
\begin{figure}
	\centering
	\includegraphics[width=0.95\linewidth]{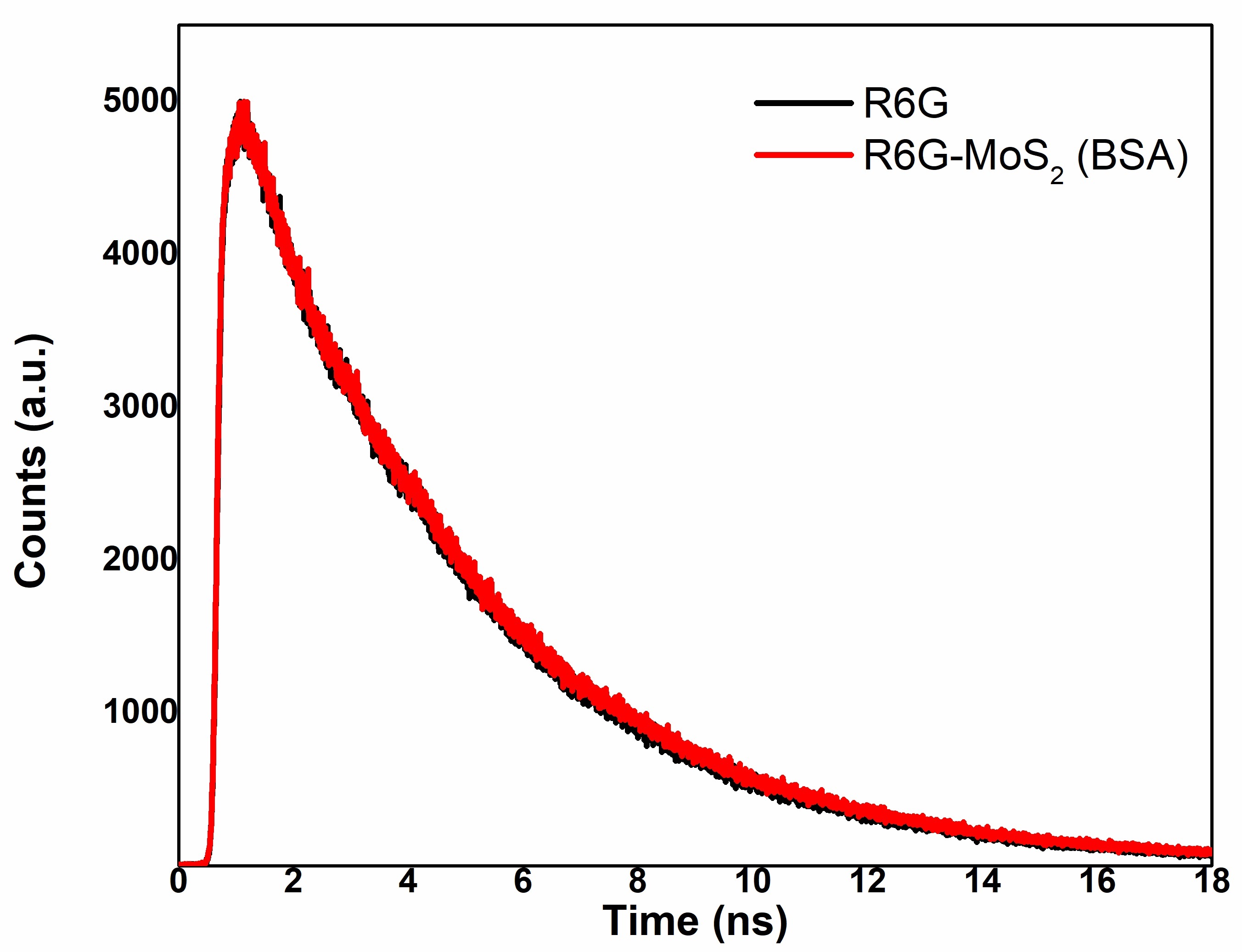} 
	\caption{Fluorescence lifetime decay curves of R6G molecules in absence and presence of MoS$_2$ nanosheets, here MoS$_2$ nanosheets are exfoliated using BSA molecules. \label{TRFS-BSA}}
\end{figure}
Figure \ref{TRFS-BSA} shows  the Fl lifetime decay curves of R6G molecules in absence and presence of MoS$_2$ nanosheets. Here the MoS$_2$ nanosheets are same as the previous case, where BSA molecules are on surface of MoS$_2$ and they act as stabilizing agent for the nanosheets. The Fl lifetime of R6G molecules is approximately 4 ns in absence of acceptors. In presence of MoS$_2$ (BSA), there is no significant change in the Fl lifetime of R6G molecules, which is due to significant distance between  R6G molecules and MoS$_2$ nanosheets. This distance is suppose to be of the order of BSA molecule, which is acting as separator between R6G and MoS$_2$. In order to shorten the distance between R6G and MoS$_2$, a new set of MoS$_2$ nanosheets has been prepared using n-methyl-2-pyrrolidone (NMP) as dispersion medium. In NMP solution, MoS$_2$ powder was dispersed and the mixture was sonicated for 20h. The supernatant  was collected and used as acceptor for energy transfer study. The Fl lifetime behaviors of R6G molecules in presence and absence of  MoS$_2$ nanosheets are shown in Figure \ref{TRFS-NMP}.        
In Figure \ref{TRFS-NMP} a significant change in the Fl lifetime of R6G molecule is observed. This indicates that there is energy transfer phenomenon taking place from R6G molecules to the MoS$_2$ nanosheets. In this case, the separation between R6G and MoS$_2$ nanosheets are within the limit of energy transfer distance (1-10 nm).  

\section{Summary}
In summary, the interaction of gold nanoclusters with the molybdenum disulfide was explored. It had been observed that the BSA,  protein molecules, stabilized gold nanoclusters had strong florescence emission around 700 nm and its emission intensity can be altered in presence of MoS$_2$ nanosheets. The florescence quenching was found to be static in the nature. It was observed when distance between donor and acceptor was short in case of R6G molecules and MoS$_2$ nanosheets the energy transfer phenomenon took place.      

\section{Acknowledgements}

The authors are thankful to AIRF and SBT, JNU for characterization techniques.

		\bibliography{main}
	


\end{document}